\documentclass[sn-mathphys,Numbered,iicol]{sn-jnl}% Math and Physical Sciences Reference Style
\usepackage{graphicx}%
\usepackage{multirow}%
\usepackage{amsmath,amssymb,amsfonts}%
\usepackage{amsthm}%
\usepackage{mathrsfs}%
\usepackage[title]{appendix}%
\usepackage{xcolor}%
\usepackage{textcomp}%
\usepackage{manyfoot}%
\usepackage{booktabs}%
\usepackage{algorithm}%
\usepackage{algorithmicx}%
\usepackage{algpseudocode}%
\usepackage{listings}%
\usepackage{float}

\theoremstyle{thmstyleone}%
\theoremstyle{thmstyletwo}%

\theoremstyle{thmstylethree}%

\begin{document}

\title[Extreme events and instantons in shell models]{Instanton-based Importance Sampling for Extreme Fluctuations in a Shell Model for Turbulent Energy Cascade}


\author*[1]{\fnm{Guilherme} \sur{Tegoni Goedert}}\email{guilherme.goedert@fgv.br}

\author[2]{\fnm{Luca} \sur{Biferale}}\email{Luca.Biferale@roma2.infn.it}

\affil*[1]{\orgdiv{School for Applied Mathematics}, \orgname{Getúlio Vargas Foundation}, \orgaddress{\street{Praia de Botafogo, 190}, \city{Rio de Janeiro}, \postcode{22250-900}, \state{Rio de Janeiro}, \country{Brazil}}}

\affil[2]{\orgdiv{Department of Physics and INFN}, \orgname{University of Rome, Tor Vergata}, \orgaddress{\street{Via della Ricerca Scientifica 1}, \city{Rome}, \postcode{00133}, \state{Lazio}, \country{Italy}}}


\abstract{Many out-of-equilibrium flows present non-Gaussian fluctuations in physically relevant observables, such as energy dissipation rate. This implies extreme fluctuations that, although rarely observed, have a significant phenomenology. Recently, path integral methods for importance sampling have emerged from formalism initially devised for quantum field theory and are being successfully applied to the Burgers equation and other fluid models. We proposed exploring the domain of application of these methods using a Shell Model, a dynamical system for turbulent energy cascade which can be numerically sampled for extreme events in an efficient manner and presents many interesting properties. We start from a validation of the instanton-based importance sampling methodology in the heat equation limit. We explored the limits of the method as non-linearity grows stronger, finding good qualitative results for small values of the leading non-linear coefficient. A worst agreement between numerical simulations of the whole systems and instanton results for estimation of the distribution's flatness is observed when increasing the nonlinear intensities.}




\maketitle

\section{Introduction}
Turbulent flow is a complex system where coherent structures are formed in time and space and interact on a wide range of spatial scales, as energy is introduced in large scales (between kilometers for atmospheric flows and centimeters for a cup of coffee) but is dissipated only at scales many orders of magnitude smaller. In between energy injection and dissipation, there is a wide range of scales where these structures, eddies, exhibit self-similar behavior and a well-recorded power-law spectrum for energy; this is called the inertial range \cite{frisch1995turbulence}.

Of particular interest to the investigation of this phenomenology are shell models for turbulent energy cascade. These models can present highly non-trivial and intermittent behavior analogous to fully developed turbulence \cite{frisch1995turbulence, ohkitani1989temporal, bohr1998dynamical, Biferale2003, Dombre1998}. Nevertheless, these dynamical systems are numerically and analytically more tractable than the Navier-Stokes equations, serving as a very interesting class of models for testing ideas and methods in fluid dynamics.

In studying out-of-equilibrium micro- and macroscopic fluids, strongly non-Gaussian and intermittent fluctuations are frequently observed. This implies that abrupt changes in fluid velocity occur more often and with larger effects than would be observed if Gaussian statistics were followed. These are often related to sparse but intense coherent structures both at large scales (e.g. atmospheric fronts, tornadoes) and small scales (influencing dissipation of energy in the atmosphere or the ocean \cite{Mahrt89, Rainville08}, or the rain formation process \cite{Falkovich07}).

However, the intrinsic rarity of these extreme events makes their statistical assessment a great challenge. To investigate extreme events in the energy cascade of classical hydrodynamic turbulence, we propose the employment of Instanton-based methods for the importance sampling of intense fluctuations and explore this methodology in Shell Models, a class of dynamical systems which mimic turbulent energy cascade while being an efficient playground for sampling of extreme event numerically.

 Instanton-based methods for importance sampling have been successfully tested for the Burgers equations \cite{Grafke2015, Margazoglou2019, apolinario2019onset,grafke2013instanton}. Furthermore, instantons as statistically dominant field configurations have been used to describe circulation intermittency in Navier-Stokes \cite{migdal2020clebsch, moriconi2022}. Here, we explore the simpler but dynamically rich setting of shell models for hydrodynamics turbulence, which have been used also as testing ground for modeling magnetohydrodynamics \cite{frick1998cascade, nigro2004nanoflares, PLUNIAN20131}, passive scalars \cite{jensen1992shell, cohen2002statistically, mitra2005dynamics} and convective turbulence \cite{brandenburg1992energy, ching2010studying}, among others.

\section{The Densniansky-Novikov shell model}
Shell models are systems of (infinitely many) coupled ordinary differential equations, modeled after the Navier-Stokes equations in Fourier space and which preserve many of this system's nonlinear properties and symmetries. One can create shell models by discretizing the Fourier space into concentric shells with radii arranged in a geometric sequence ($k_n = k_0 h^n$), and keeping only one variable $u_n(t)$ -or a few variables-representing each shell. This is a further simplification with respect to projecting the equation into a logarithmic lattice \cite{campolina2018chaotic}, and this greatly decreases the number of degrees of freedom required to simulate solutions (to the order of the logarithm of the Reynolds number, rather than a power of it). 

The real-valued Desniansky-Novikov (D.N.) shell model, generalized to extend the Obukhov model and with a  stochastic forcing, reads \cite{Ditlevsen, Dombre1998}
\begin{equation}
\begin{split}
\label{DNS_DN_eq:real_model}
\dfrac{d u_n}{dt} =& \  G_n[u] + f_n, \ \  \left\langle f_n (t) f_m (t') \right\rangle = \chi_{nm} \delta (t - t')\\
G_n[u] =& \ c_1 (k_n u_{n-1}^2 - k_{n+1} u_n u_{n+1}) \\+ &\ c_2 (k_n u_{n-1} u_n - k_{n+1} u_{n+1}^2) - \nu k_n^2 u_n,
\end{split}
\end{equation}
\noindent where $c_1$, $c_2$ and $\nu$ are real constant parameters, with $\nu \geq 0$ corresponding to the viscosity and $\chi_{nm}$ is the correlation between the forcing applied to different shells. Although this is an infinite-dimensional dynamical system, in practice we need to truncate the shell space to $n \in \left\lbrace 0, 1, ..., N \right\rbrace$ while imposing boundary conditions $u_{-1} = u_{N+1} = 0$ and choosing $N$ large enough to include integral, inertial, and viscous scales \cite{frisch1995turbulence}. This model was constructed to conserve total energy $E = \frac{1}{2}\sum_n u_n^2$ in inviscid and unforced settings.

\begin{figure}
	\centering
	\includegraphics[width = 0.45\textwidth]{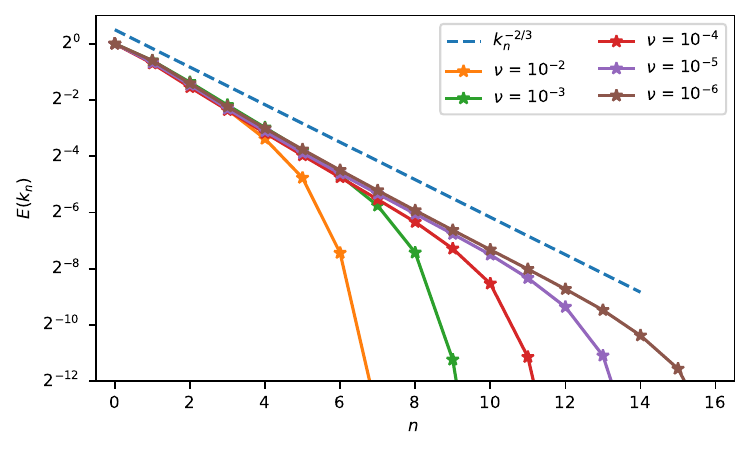}
	\caption{Mean energy values per shell $E(k_n)$ in the D.N. Shell Model \eqref{DNS_DN_eq:real_model}. Parameters $c_1 = 1$, $c_2 = 0$ regulate the energy transfer between scales, while we vary the values of viscosity $\nu$. }
    \label{fig:shell_models_Ek}
\end{figure}

Among the key similarities between Shell Models and fully developed turbulence, is the existence of a turbulent energy cascade ``à la Kolmogorov". An example of energy spectrum is shown in Figure \ref{fig:shell_models_Ek} for the Desniansky-Novikov shell model \eqref{DNS_DN_eq:real_model} where we verify that $E(k_n) = \langle |u_n|^2 \rangle \propto k_n^{-2/3}$, which is the equivalent of the $-5/3$ power-law spectrum in three dimensional isotropic turbulence considering the extra factor included by the thickness of each shell. Here and hereafter time average is denoted with $\langle \cdot \rangle$.  Moreover, shell models also present highly intermittent bursts in its energy dissipation rate $\epsilon = \nu \sum_n k_n^2 u_n^2$, with many episodes an order of magnitude or more higher than the average value. Nonetheless, these rare events are a lot easier to sample in Shell Models than in the Navier-Stokes equation highlighting the usefulness of these simpler models, specially in the study of extreme events. 

\begin{figure*}
    \centering
    \includegraphics[width=0.95\textwidth]{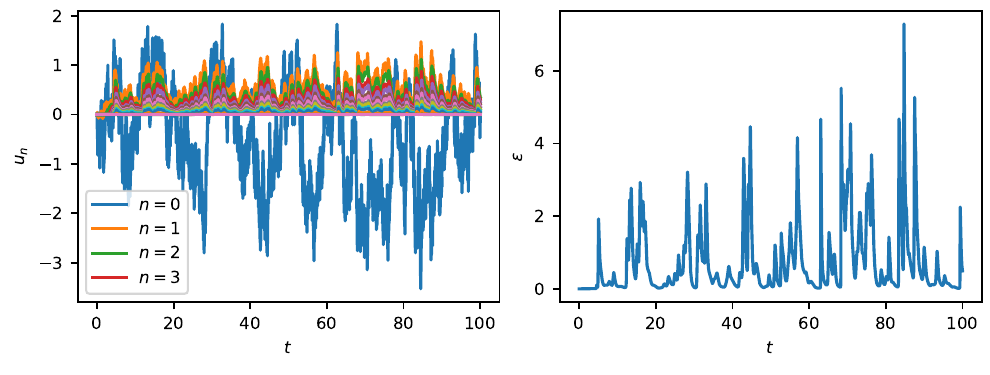}
    \caption{Solution for the stochastically forced D.N. shell model \eqref{DNS_DN_eq:real_model} on the left and corresponding time series for the energy dissipation rate $\epsilon = \nu \sum k_n^2 u_n^2$ on the right; parameters used: $c_1 = c_2 = 0.5$, $\nu=10^{-4}$ and $\chi_{nm} = k_n^{-6}\delta_{nm}$.}
    \label{fig:stochastic_solution}
\end{figure*}

\begin{figure}
	\centering
	\includegraphics[width = 0.45\textwidth]{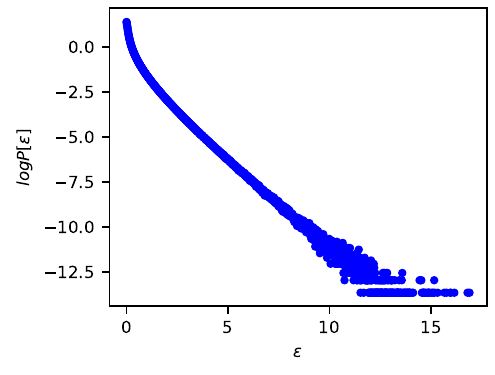}
	\caption{PDF of energy dissipation rate $\epsilon$  for a large dataset with $10^8$ samples,  in the weakly linear setting explored for validation of our methodology ($c_1 = 10^{-3}$, $\nu=10^{-2}$, $c_2 = 0$ and $\chi_{nm} = k_n^{-6}\delta_{nm}$). The average value $\langle \epsilon \rangle = 0.5166$,  }
    \label{fig:eps_pdf}
\end{figure}

The D.N. shell model was chosen for this study due to its simplicity as well as its different asymptotic behaviors when changing the control parameters $c_1$ and $c_2$ \cite{Dombre1998, Ditlevsen}. 
%energy moves only from smaller to larger shells (equivalently, from large to small physical scales) while it moves in reverse for $c_2 \neq 0$ and $c_1 = 0$. Such a clear example of a physically relevant system where the balance between the two possible directions of the energy cascade can be controlled by one parameter is rare and valuable when testing the domain of application of new methodologies since we can first isolate two types of dynamical behavior. 
In the absence of stochastic forcing, the solutions of this shell model tend to fixed-point attractors compatible with the power-law scaling found in the energy spectrum. As soon as a stochastic forcing is introduced, the system is driven out of equilibrium  while still preserving the scaling predicted by the energy spectrum (see Figure \ref{fig:stochastic_solution}). At the same time, intermittent but large fluctuations appear in the high wavenumber shells, translating into bursts in the energy dissipation rate much larger than its average value which are analogous to those observed in turbulent flows. Even in a weakly nonlinear setting (see later Section \ref{sec:results}), the distribution of the energy dissipation rate showed heavy tails as seen in Figure \ref{fig:eps_pdf}.

\section{Instanton calculation and importance sampling}
Strong non-Gaussian and intermittent fluctuations are observed in many out-of-equilibrium micro and macroscopic fluids, implying that abrupt changes in fluid velocity happen more often and with larger effects than would be observed if Gaussian statistics were followed. These large gradients are often related to sparse but intense coherent structures at both large scales (e.g., atmospheric fronts, tornadoes) and small scales (influencing dissipation of energy in the atmosphere or the ocean \cite{Mahrt89, Rainville08}, or the rain formation process \cite{Falkovich07}).

Despite the importance of such phenomena, their rarity leads to exceeding difficulties in their observation, in the real world or in simulations. Therefore, it is key to have numerical tools capable of quantitatively assessing the statistical weight of these coherent or quasicoherent spatio-temporal instanton-type solutions on the statistical ensemble.

% Overview of methods being studied
Only very recently, new numerical techniques based on constrained Monte Carlo approaches emerging from the lattice QCD community have been proposed to follow this strategy in simple one-dimensional stochastic partial differential equations \cite{Duben2008, Farazmand2017, Margazoglou2019, Mesterhazy2011, Schafer:2004xa}. 

\subsection{Path integral formulation of stochastic differential equations}
These techniques are mainly based on path integral  Martin-Siggia-Rose-Janssen-de Dominicis (MSRJD) formalism \cite{Martin1973, Janssen1976, dominicis1976technics}. The general set up can be described by a  differential equation for a field $u(x,t)$, forced by some normally distributed random noise $f$ which is white in time and has some desired correlation $\chi$, 
\begin{equation}
\begin{split}
\label{eq:dynamics}
    &\dot{u} = N[u] + f(t), \\ 
    &\left\langle f(t), f(t') \right\rangle = \chi \delta(t - t').
\end{split}
\end{equation}
\noindent In the context of shell models \eqref{DNS_DN_eq:real_model}, $u = [u_0, u_1, ...]$, $N[u] = [G_0[u], G_1[u],...]$ and the stochastic forcing $f$ is also defined element-wise for the discrete shell space $k_n$ as $f(t) = [f_0, f_1, ...]$ with $\left\langle f_n (t) f_m (t') \right\rangle = \chi_{n,m} \delta (t - t')$. 
The MSRJD formalism allows for the direct evaluation of the mean value of a desired observable $\mathcal{O}[u]$ by using a  path integral written as the Onsager-Machlup functional \cite{Onsager1953, Grafke2015-2}:
\begin{equation}
\label{eq:Onsager_Machlup}
\begin{split}
    &\left\langle \mathcal{O}[u] \right\rangle = \int \mathcal{D}u \mathcal{O}[u]J[u] e^{-S_\mathcal{L}[u,\dot{u}]},\\
    &S_\mathcal{L}[u,\dot{u}] = \dfrac{1}{2} \int dt \left\langle \dot{u} - N[u], \chi^{-1}(\dot{u} - N[u]) \right\rangle,
\end{split}
\end{equation}
\noindent {where $J[u]$ is the Jacobian for the transformation between a noise realisation and velocity $f \rightarrow u$}, $\mathcal{D}u$ is the element over which functional integration is to be performed and $S_\mathcal{L}[u,\dot{u}]$ is the Lagrangian action defined by this formalism for the system \eqref{eq:dynamics}.

We perform the Hubbard-Stratonovich transformation \cite{stratonovich1957method, hubbard1959calculation} on \eqref{eq:Onsager_Machlup}, therefore, defining new momenta variables $p(x,t)$ and circumventing the dependency on the inverse correlator $\chi^{-1}$. The resulting path integral is known as the MSRJD response functional 
\begin{equation}
\label{eq:Onsager_Machlup_Hamiltonian}
\begin{split}
    &\left\langle \mathcal{O}[u] \right\rangle = \int \mathcal{D}u \mathcal{D}p \mathcal{O}[u]J[u] e^{-S_\mathcal{H}[u,p]},\\
    &S_\mathcal{H}[u,p] = \int dt \left[ \left\langle p, \dot{u} - N[u] \right\rangle - \dfrac{1}{2} \left\langle p , \chi p \right\rangle \right].
\end{split}
\end{equation}
In the general context of shell models,  of particular interest is the use of observable that constrains certain  functionals $O[u(t = 0)]$ of the fields to desired values {
\begin{equation}
    \label{eq:observable}
    \mathcal{O}[u] = \delta (O[u(t = 0)] - a).
\end{equation} 
\noindent The time at which the observable \eqref{eq:observable} is evaluated is arbitrary when dealing with autonomous systems and we adopt the convention where the observation time is shifted to the origin and the system is integrated over negative values of $t$.} In this case, the system may be rewritten in terms of a constrained action $S_\lambda[u,p]$
\begin{equation}
\label{eq:Onsager_Machlup_constrained}
\begin{split}
    &\left\langle \mathcal{O}[u] \right\rangle = \int \mathcal{D}u \mathcal{D}p \dfrac{1}{2\pi} \int_{-\infty}^{\infty} d\lambda e^{-S_\lambda[u,p]},\\
    &S_\lambda [u,p] = \lambda \left( O[u] - a \right) + S_\mathcal{H}[u,p].
\end{split}
\end{equation}
By redefining the problem from a stochastic differential system to a path integral formulation, one paves the way to the use of completely different methodological tools. Let us notice that  Hybrid Monte Carlo (HMC) \cite{DUANE1987216, luscher2010computational, MCMHandbook} has also been successfully implemented via modification of the Onsager-Machlup functional to define an artificial Hamiltonian whose evolution is used to find candidates for states in a Metropolis–Hastings algorithm; this method is explored in \cite{Margazoglou2019} for the stochastically forced Burgers equations.

\subsection{Instanton equations for shell models}
Instantons are field configurations that are extrema of the action, that is, they are the largest contributors to the evaluation of the mean value of an observable in the path integral formulation. Therefore, they can be found as solutions to the equations obtained via a saddle-point approximation. 
Here, we are interested to realizations where a given function attains  desired value. An instanton is therefore found as a solution to the system of equations derived through a saddle-point approximation of the action in \eqref{eq:Onsager_Machlup_constrained},
\begin{equation}
\label{eq:saddle_conditions}
    \dfrac{\delta S_\lambda}{\delta u} = 0 , \quad \dfrac{\delta S_\lambda}{\delta p} = 0.
\end{equation}
\noindent for a system of the form \eqref{eq:dynamics}, the conditions  \eqref{eq:saddle_conditions} lead to a coupled set of equations between the original field and the momenta: 
\begin{equation}
\label{eq:instanton_dynamics}
\begin{split}
    \dot{u} &= N[u, x] + \chi \ast p,\\
    \dot{p} &= -\left( \nabla_u N[u] \right)^T p + \lambda \nabla_u O[u(t)] \delta (t),
\end{split}
\end{equation}
where $\ast$ stands for the convolution product and {$\delta$ is a Dirac delta that is nonzero at the observation time defined in \eqref{eq:observable}. The idea is that we want to observe an extreme fluctuation at this observation time, giving it enough time to develop \cite{grafke2013instanton}. Therefore, this system is defined on $t \in (-T, 0]$ where $T>0$ is chosen large enough to satisfy condition $u$, $p \rightarrow 0$ as $t \rightarrow -\infty$. We observe that due to the Dirac delta in the second equation of \eqref{eq:instanton_dynamics}, the term $\lambda \nabla_u O[u(t)]$ is effective only at the observation time $t = 0$ at the right extreme of the time interval; this term can therefore be removed from the equation and considered as a boundary condition.} 
The MSRJ response functional for a generic shell model is: 
\begin{equation}
\begin{split}
\label{eq:action_shell_models}
    S_\mathcal{H} [u,p] = \int dt \left( \sum_{m = 0}^N p_m (\dot{u}_m - G_m[u]) \right. \\ \left. - \dfrac{1}{2} \sum_{m = 0}^N \sum_{l = 0}^N p_m \chi_{ml} p_l \right).
\end{split}
\end{equation}

\noindent The instanton equations then become ($s$ being the number of shells between the farthest interacting neighbouring shell in the model)
\begin{equation}
%\begin{split}
\label{eq:instanton_shell_models}
    \dot{u}_n = G_n[u] + \sum_{l = 0}^N \chi_{nl} p_l \ , \quad \dot{p}_n + \sum_{m = n - s}^{n+s} p_m \dfrac{\partial G_m}{\partial u_n} = 0, 
%\end{split}
\end{equation}
\noindent {defined on $t \in (-T, 0]$ which leads to the boundary condition $p_n(0) = \lambda \nabla_u O[u(0)]$. We choose $T>0$ large enough to satisfy the condition $u$, $p \rightarrow 0$ on the left boundary.} 
For the real-valued Desniansky-Novikov model \eqref{DNS_DN_eq:real_model}, the instanton equations become
\begin{equation}
\label{eq:instanton_shell_model_dn}
\begin{split}
    \dot{u}_n + \nu k_n^2 u_n&= c_1 k_n \left( u_{n-1}^2 - hu_n u_{n+1} \right) \\ &+ c_2 k_n \left( u_{n-1} u_n - h u_{n+1}^2 \right) + \sum_{l = 0}^N \chi_{nl} p_l, \\
    \dot{p}_n - \nu k_n^2 p_n&= -c_1 ( 2k_{n+1} u_{n}p_{n+1} - k_{n+1} u_{n+1} p_n \\ &  - k_n u_{n-1} p_{n-1} ) - c_2 ( k_{n+1} u_{n+1} p_{n+1} \\  & + k_{n} u_{n-1} p_n - 2 k_n u_n p_{n-1} ),
\end{split}
\end{equation}
with the extra constraints $p_n(0) = -i \lambda \nabla_u O[u(0)]$ and $u, p \rightarrow 0$ as $t \rightarrow -\infty$. 

\subsection{Chernykh-Stepanov iterative method} \label{subsec:iterative_method}
\begin{figure}[ht]
    \centering
    \includegraphics[width=0.5\textwidth]{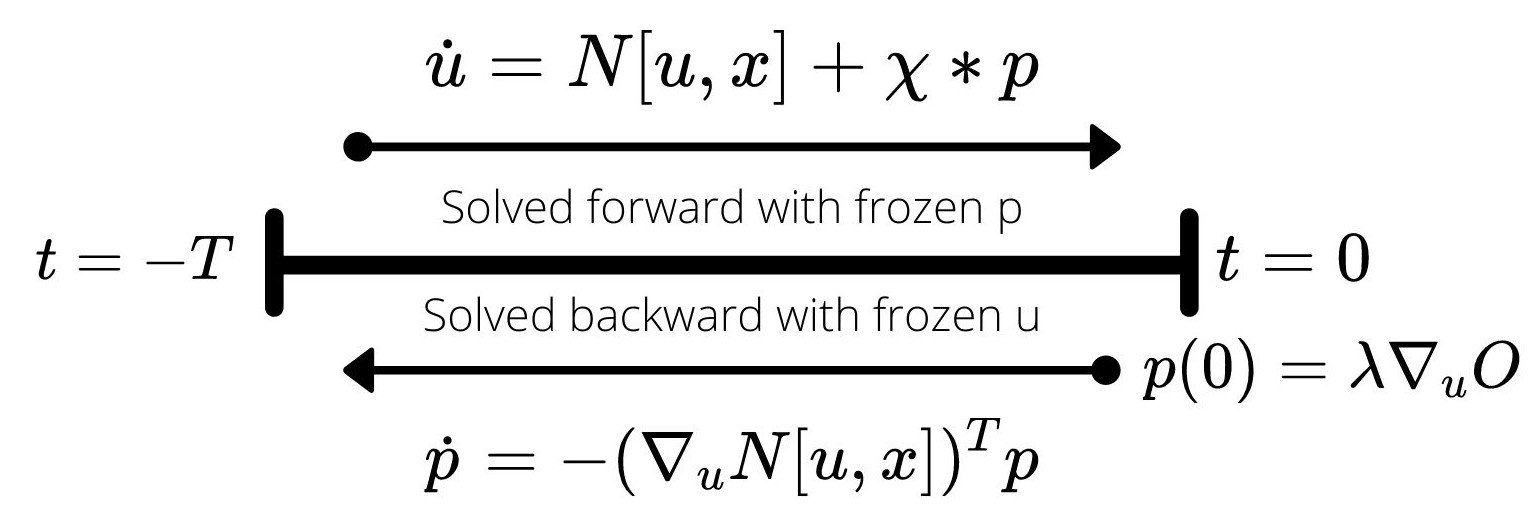}
    \caption{Graphic representation of the Chernykh-Stepanov iterative method.}
    \label{Ch2_fig:instanton_scheme}
\end{figure}

Though simple in nature, instanton equations generally pose an interesting numerical challenge: if $N[u]$ contains a linear dissipative term, as is usually the case in viscous systems, the equations for $u$ and $p$ are unstable if integrated together in the same direction of time (see the {different signs of the viscous terms in system \eqref{eq:instanton_shell_model_dn}) \cite{grafke2013instanton, Grafke2015-2}}.
As such, we need to employ a specialized iterative method that splits the dynamics of each field and solves each along its stable direction in time, as described in Figure \ref{Ch2_fig:instanton_scheme} \cite{Stepanov, grafke2013instanton, Grafke2015-2, Grafke2015, Ebener2018}. Note that the dynamics \eqref{eq:instanton_shell_model_dn} of one field includes a coupling term with the other; in order to evaluate this coupling, we need to solve both equations at the same instant of time, and we use the approximations from the last iteration of the method at the same instant to evaluate the coupling terms during integration. Note that the term accompanying the Dirac delta in the evolution of $p$ can be considered a simple initial condition $p(0) = -i \lambda \nabla_u O[u]$. Observe that the constrained value $a$ of the functional $O[u]$ does not take part in the dynamics; the instanton configuration for a given dynamics and functional depends only on the parameter $\lambda$ that arises from the Fourier transform of {the constraint \eqref{eq:observable} in order to incorporate it into the  action \eqref{eq:Onsager_Machlup_constrained}}. The constraint value is only used in the calculation of the action $S_\lambda$, which is used to calculate the PDF for the observable studied {up to a prefactor as $\exp(-S_\lambda)$.}

\section{Solving instanton equations {in shell models}}

\subsection{Locality of observables and forces}
Observables and forces play a key role in the formulation of instanton equations and even more so in their solution. It is the spatial correlator $\chi$ of the forces that couple the dynamics of $u$ to the configuration of the auxiliary variable $p$. Also, it is the u-gradient of the constraint functional $O[u]$ that injects energy into the system by serving as the initial condition for $p$ as it is solved backward in time. However, not all choices of force and observable are compatible {in order to develop non trivial instanton solutions}.

In order to initiate the Chernykh-Stepanov iterative scheme, we begin with the fields at rest with $u(t) = p(t) = 0$ everywhere except the right boundary where $\lambda \nabla_u O[u(0)]$ is enforced for a fixed value of $\lambda$. The system is excited by the right boundary condition and energy it introduces is initially propagated due to the term $\sum_l \chi_{nl} p_l$. Let us suppose that an observable local to shell $n_o$ is chosen {(i.e., the observable is a function only of the component $n_o$ of $u$)}, while a different shell $n_f$ is the only one forced. Since $G_n[u]$ {is composed by a linear viscous term and a sum of bilinear terms of shifts of $u$, it is easy to see that in the first iteration of the Chernykh-Stepanov method, only $p_{n_o} \neq 0$. However, if a force is local and $n_o \neq n_f$, then $\dot{u}_n \equiv 0$ for all $n$ in \eqref{eq:instanton_shell_models} and this field remains at rest. Looking at the equation for $\dot{p}_n$, either in its generic form \eqref{eq:instanton_shell_models} or in its reduction to the D.N. shell model \eqref{eq:instanton_shell_model_dn}, we see that all but the viscous term depend linearly on shell velocities $u_n$ and will therefore cancel out if $u$ remains at rest. In this way, the Chernykh-Stepanov will converge to a trivial instanton solution.}  

Therefore,  one must take care to choose at least global forces or observables, if not both. Moreover, local forces present ill-conditioned inverse correlators.  To sidestep both issues, we studied in detail how the system responds to power law forcing correlations $$\chi_{nm} = k_n^{x}\delta_{nm}$$ and choose a value of the exponent $x$ small enough  ($x\leq -6$) such as to mimic an {\it almost local} forcing as in realistic three dimensional turbulent flows.  On the opposite, for large exponent values, the system is fully driven by the Gaussian forcing at all scales; this constitutes another interesting limiting scenario for validation.

\subsection{Instanton solutions}
\begin{figure*}
    \centering
    \includegraphics[width=0.95\textwidth]{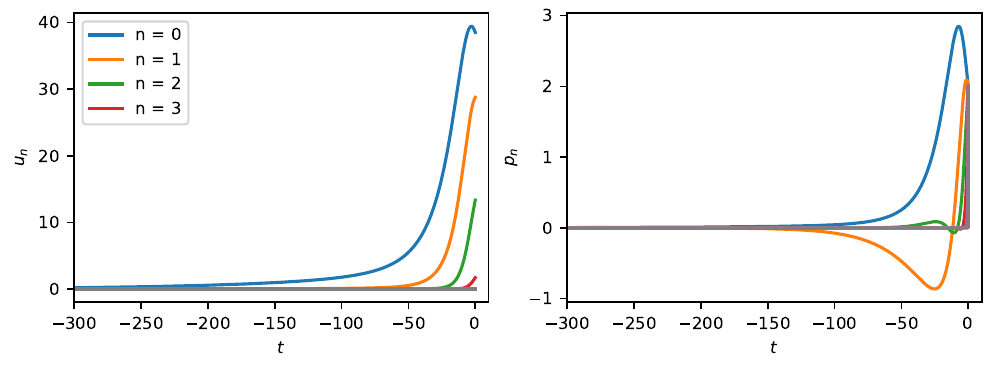}
    \caption{Instanton solution for \eqref{eq:instanton_shell_model_dn} for $\lambda = 2$,  $O = \sum_n u_n = 82.34$; $c_1 = 10^{-3}$, $\nu=10^{-2}$, $c_2 = 0$ and $\chi_{nm} = k_n^{-6}\delta_{nm}$.}
    \label{Ch2_fig:instanton_solution}
\end{figure*}

\begin{figure*}
    \centering
    \includegraphics[width=0.95\textwidth]{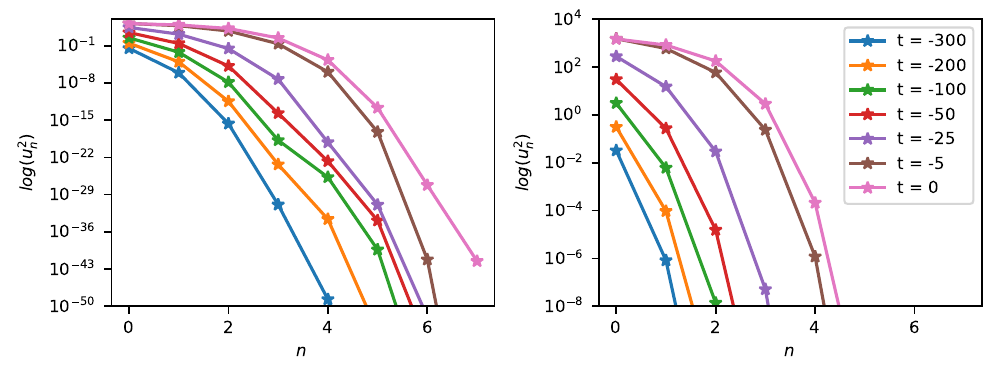}
    \caption{Squared velocities of each shell in semi-log scale for an instanton approximation as it develops over different time slices. Parameters used: $\lambda = 2$,  $O = \sum_n u_n = 82.34$; $c_1 = 10^{-3}$, $\nu=10^{-2}$, $c_2 = 0$ and $\chi_{nm} = k_n^{-6}\delta_{nm}$.}
    \label{Ch2_fig:instanton_solution_Ek}
\end{figure*}

In this first work we focus on results about the simplest linear global observable 
\begin{equation}
    O[u] = \sum_n u_n.
\end{equation}
For each multiplier value $\lambda$, we seek a numerical approximation to the corresponding instanton using the Chernykh-Stepanov iterative method. To this end, we start with null fields $u$ and $p$ on a uniform grid on the time interval $[-T, 0]$.  Then, we perform the iterative method described in Subsection \ref{subsec:iterative_method} while employing a suitable integrator; here, we used the two-step Adam-Bashforth method. {Figure \ref{Ch2_fig:instanton_solution} depicts an instanton solution obtained in this way for $\lambda = 2$, having converged after 175 interations. We observe that as $t$ increases, more shells are excited and contribute to the evaluation of the observable for which the instanton was defined; this can be seen in the evolving time slices of the energy spectrum shown in Figure \ref{Ch2_fig:instanton_solution_Ek}.}

\begin{figure}
    \centering
    \includegraphics[width=0.5\textwidth]{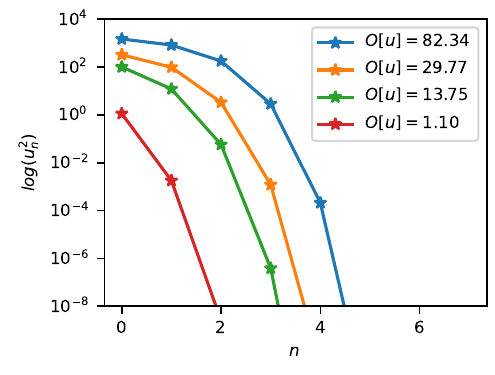}
    \caption{Squared velocities of different shells for the final fully developed instantons ($t = 0$) obtained  for various values of $\lambda$; curves are labelled according to the corresponding value of the observable $O = \sum_n u_n$. Parameters used: $c_1 = 10^{-3}$, $c_2 = 0$, $\nu=10^{-2}$ and $\chi_{nm} = k_n^{-6}\delta_{nm}$.}
    \label{Ch2_fig:instanton_solution_Ek_lamb}
\end{figure}

Having solved the instanton equations for a given value of $\lambda$, we evaluate the observable and the action mapped to this value. By collecting these values for a sufficient range of $\lambda$'s, we can evaluate the PDF and the mean value of the observable using equations \eqref{eq:Onsager_Machlup_constrained}.

The validation of the methodology and implementation code was performed by observing that in the fully linear case ($c_1 = c_2 = 0$) each equation of the shell becomes uncoupled and becomes an Ornstein-Uhlenbeck process. In this setting, we can compare the distributions of an observable obtained from direct numerical simulation (DNS) and instanton calculations, we also have analytical results for the average and variance of the distributions of each shell velocity. 

\subsection{Tail distributions for extreme fluctuations in the shell model} \label{sec:results}
As one can see from Figure \ref{Ch2_fig:instanton_solution_Ek_lamb}, instantons leading to high values of $O$ are characterized by a non trivial energy distribution among many shells, supporting the idea that the system is developing non trivial multi-scale correlation to match the constrain. The PDF for an observable is computed up to a prefactor from evaluation of the constrained action $S_\lambda$ for a range of values of $\lambda$ as $\exp(-S_\lambda)$. In order to understand how well this PDF constructed only from instantons approximates the distribution of the observable $O$ originally obtained from direct numerical simulation of the shell model, we systematically explore the domain of application of the methodology, starting with the fully linear setting we used for validation and gradually increasing the values of the coefficient $c_1$. 

In Figures \ref{fig:inst_DN_c1_1E-4_PDF} and \ref{fig:inst_DN_c1_1E-3_PDF} we show the PDFs of the observable compared with the one from the numerical simulation (DNS) both in lin-lin and log-lin scale for two  values of non-linearity $c_1= 10^{-4}$ and $10^{-3}$, respectively. As one can see, the agreement is pretty good, not only for very intense values of the observable. By increasing the non-linearity, the agreement deteriorate, as shown in Figure \ref{fig:flatness} by the calculation of the {excess kurtosis: 
$$F =\frac{\langle (O - \langle O\rangle )^4 \rangle }{\langle (O - \langle O\rangle )^2 \rangle^2 }-3.$$  despite this, the tail of the distributions still shows good agreement also for this relatively larger values of nonlinearities, as shown in Figure \ref{fig:inst_DN_c1_1E-3_PDF}. By increasing even further 
$c_1$, the agreement between the calculations based only on the contributions from instantons and the full DNS is worsening. 
%\begin{figure}[H]
%    \centering
%    \includegraphics[width=0.45\textwidth]{figures/nu_1E-02_c1_1E-03_c2_0E+00_x_-3E+00_DNS_spectra.pdf}
%    \caption{Energy spectrum of DNS solution; $c_1 = 10^{-3}$, $\nu=10^{-2}$, $c_2 = 0$ and $f_n = k_n^{-3}$.}
%    \label{fig:inst_DN_c1_1E-3_spectrum}
%\end{figure}

\begin{figure*}
    \centering
    \includegraphics[width=0.95\textwidth]{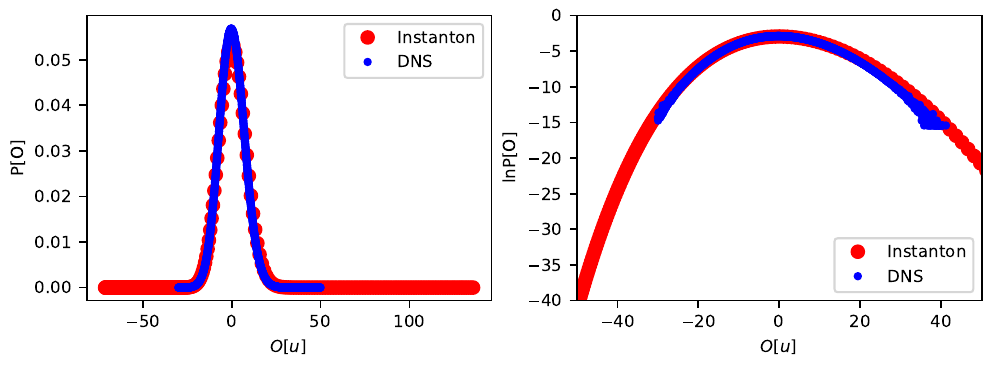}
    \caption{Comparison of PDFs for the observable $O = \sum_n u_n$ obtained from DNS and instanton calculations, linear scale to the left and log scale in the y-axis to the right. Parameters used: $c_1 = 10^{-4}$, $c_2 = 0$, $\nu=10^{-2}$  and $\chi_{nm} = k_n^{-6}\delta_{nm}$.}
    \label{fig:inst_DN_c1_1E-4_PDF}
\end{figure*}

\begin{figure*}
    \centering
    \includegraphics[width=0.95\textwidth]{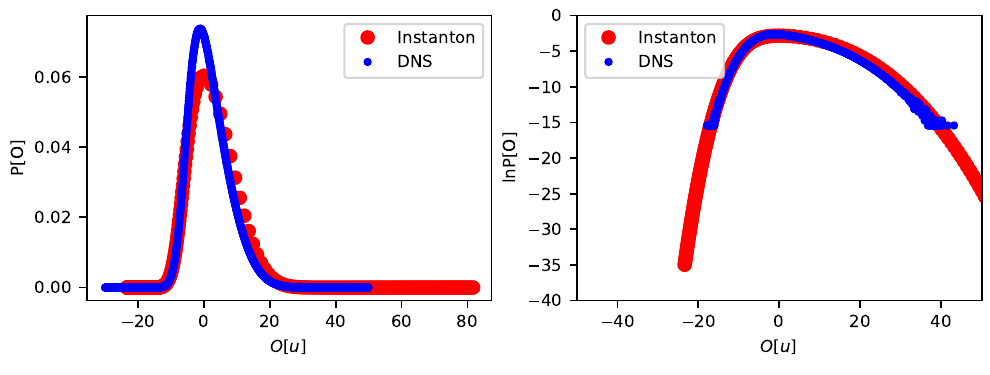}
    \caption{Comparison of PDFs for the observable $O = \sum_n u_n$ obtained from DNS and instanton calculations, linear scale to the left and log scale in the y-axis to the right. Parameters used: $c_1 = 10^{-3}$, $c_2 = 0$, $\nu=10^{-2}$  and $\chi_{nm} = k_n^{-6}\delta_{nm}$.}
    \label{fig:inst_DN_c1_1E-3_PDF}
\end{figure*}

\begin{figure}
    \centering
    \includegraphics[width=0.45\textwidth]{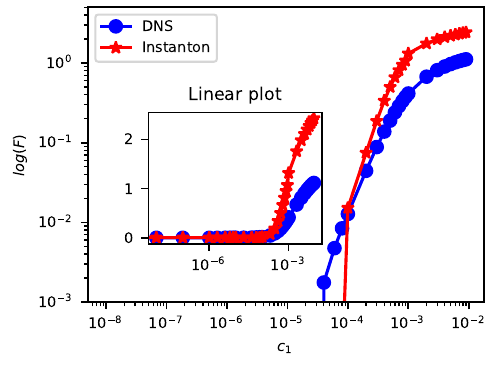}
    \caption{Comparison of excess kurtosis for PDFs of the observable $O = \sum_n u_n$ for varying values of $c_1$ with $\nu=10^{-2}$, $c_2 = 0$ and $\chi_{nm} = k_n^{-6}\delta_{nm}$.}
    \label{fig:flatness}
\end{figure}

\section{Discussion and conclusions}
We conclude that the shell models, and especially the Desniansky-Novikov shell model, are excellent candidates for testing the importance sampling methodology,
allowing for extensive sampling even of rare events with relative simplicity if compared for full Navier-Stokes equations or other fluid models. Starting from the 'Heat equations' limit provides excellent validation ground for the new methods, as the models reduce to a sequence of Ornstein-Uhlenbeck processes, where analytical and numerical results are readily available.
The MSRJD formalism was applied to Shell Models in general, and instanton equations were successfully defined and solved using the Chernykh-Stepanov iterative method.
We observed a good agreement between DNS and instanton-based PDFs in the regime of weak nonlinearity where dynamics in all shells is still mainly dominated by dissipative terms, even as the PDFs deviate from Gaussianity due to ever increasing asymmetry in the distributions.
However, as soon as nonlinearity becomes relevant in the first few shells and outliers increase in magnitude in the energy dissipation rate, we observe a difference between DNS and instanton-based results. Instantons predict larger kurtosis values and heavier positive tails.
We expect that the results can be improved by enhancing the method to take into account second-order fluctuations around the instantons \cite{Schorlepp_2021, moriconi2014velocity, apolinario2019instantons}, here obtained by considering only saddle-point approximations in the functional series expansion of the action. Improvements in similar settings have been found by \cite{apolinario2019onset} for Burgers turbulence. Work in this direction will be reported elsewhere. 
A very important question is also connected to the set of observable used to constraint the action, and more work must be done to explore instantons for small-scales quantities as the gradients $\sum_n k_n u_n$ or the energy dissipation $\sum_n k_n^2 u_n^2$.

Extensions to D.N. models with complex variables, which presents more interesting dynamics with chaotic attractors, would also be of interest. We note that the formalism presented here is defined for shell models in general by simply changing the functional $G_n[u]$ in the action \eqref{eq:action_shell_models} and instanton equations \eqref{eq:instanton_shell_models}. Instanton-based importance sampling in the GOY and SABRA shell models for hydrodynamic turbulence \cite{yamada1988goy, lvov1998sabra, Ditlevsen} as well as in shell models modeling magnetohydrodynamics \cite{frick1998cascade, nigro2004nanoflares, PLUNIAN20131}, passive scalars \cite{jensen1992shell, cohen2002statistically, mitra2005dynamics} and convective turbulence \cite{brandenburg1992energy, ching2010studying} can also be investigated. 

\backmatter

\bmhead{Acknowledgments}
We warmly thank Prof. Rainer Grauer and Timo Schorlepp (Ruhr-University Bochum) for the valuable discussions and advice. {We also thank Prof. Karl Jansen (DESY-Zeuthen) for his valuable advice and for access to the Warp cluster where simulations were performed.} This work has received funding from the European Union Horizon 2020 research and innovation programme under the Marie Skłodowska-Curie grant agreement No 765048 and by the European Research Council (ERC) under the European Union's Horizon 2020 research and innovation programme (grant agreement No. 882340). %{\color{red} EVERYONE, PLEASE SEND ME ANY ADDITION}

%\section*{Declarations}
%\subsection*{Funding}
%GTG has received funding from the European Union Horizon 2020 research and innovation programme under the Marie Skłodowska-Curie grant agreement No 765048. {\color{red} EVERYONE, PLEASE SEND ME ANY ADDITION}

%\subsection*{Conflict of interest/Competing interests}
%The authors have declared that there are no competing interests.

%\subsection*{Ethics approval }
%Not applicable.

%\subsection*{Consent to participate}
%Not applicable.

%\subsection*{Consent for publication}
%The authors consent to the publication of this work after the review process.

%\subsection*{Availability of data and materials}
%Not applicable.

%\subsection*{Code availability}
%{\color{red}TODO}

%\subsection*{Authors' contributions}
%{\color{red}TODO}



\begin{thebibliography}{45}
\ifx \bisbn   \undefined \def \bisbn  #1{ISBN #1}\fi
\ifx \binits  \undefined \def \binits#1{#1}\fi
\ifx \bauthor  \undefined \def \bauthor#1{#1}\fi
\ifx \batitle  \undefined \def \batitle#1{#1}\fi
\ifx \bjtitle  \undefined \def \bjtitle#1{#1}\fi
\ifx \bvolume  \undefined \def \bvolume#1{\textbf{#1}}\fi
\ifx \byear  \undefined \def \byear#1{#1}\fi
\ifx \bissue  \undefined \def \bissue#1{#1}\fi
\ifx \bfpage  \undefined \def \bfpage#1{#1}\fi
\ifx \blpage  \undefined \def \blpage #1{#1}\fi
\ifx \burl  \undefined \def \burl#1{\textsf{#1}}\fi
\ifx \doiurl  \undefined \def \doiurl#1{\url{https://doi.org/#1}}\fi
\ifx \betal  \undefined \def \betal{\textit{et al.}}\fi
\ifx \binstitute  \undefined \def \binstitute#1{#1}\fi
\ifx \binstitutionaled  \undefined \def \binstitutionaled#1{#1}\fi
\ifx \bctitle  \undefined \def \bctitle#1{#1}\fi
\ifx \beditor  \undefined \def \beditor#1{#1}\fi
\ifx \bpublisher  \undefined \def \bpublisher#1{#1}\fi
\ifx \bbtitle  \undefined \def \bbtitle#1{#1}\fi
\ifx \bedition  \undefined \def \bedition#1{#1}\fi
\ifx \bseriesno  \undefined \def \bseriesno#1{#1}\fi
\ifx \blocation  \undefined \def \blocation#1{#1}\fi
\ifx \bsertitle  \undefined \def \bsertitle#1{#1}\fi
\ifx \bsnm \undefined \def \bsnm#1{#1}\fi
\ifx \bsuffix \undefined \def \bsuffix#1{#1}\fi
\ifx \bparticle \undefined \def \bparticle#1{#1}\fi
\ifx \barticle \undefined \def \barticle#1{#1}\fi
\bibcommenthead
\ifx \bconfdate \undefined \def \bconfdate #1{#1}\fi
\ifx \botherref \undefined \def \botherref #1{#1}\fi
\ifx \url \undefined \def \url#1{\textsf{#1}}\fi
\ifx \bchapter \undefined \def \bchapter#1{#1}\fi
\ifx \bbook \undefined \def \bbook#1{#1}\fi
\ifx \bcomment \undefined \def \bcomment#1{#1}\fi
\ifx \oauthor \undefined \def \oauthor#1{#1}\fi
\ifx \citeauthoryear \undefined \def \citeauthoryear#1{#1}\fi
\ifx \endbibitem  \undefined \def \endbibitem {}\fi
\ifx \bconflocation  \undefined \def \bconflocation#1{#1}\fi
\ifx \arxivurl  \undefined \def \arxivurl#1{\textsf{#1}}\fi
\csname PreBibitemsHook\endcsname

%%% 1
\bibitem[\protect\citeauthoryear{Frisch}{1995}]{frisch1995turbulence}
\begin{bbook}
\bauthor{\bsnm{Frisch}, \binits{U.}}:
\bbtitle{Turbulence: the Legacy of AN Kolmogorov}.
\bpublisher{Cambridge University Press},
\blocation{New York}
(\byear{1995})
\end{bbook}
\endbibitem

%%% 2
\bibitem[\protect\citeauthoryear{Ohkitani and
  Yamada}{1989}]{ohkitani1989temporal}
\begin{barticle}
\bauthor{\bsnm{Ohkitani}, \binits{K.}},
\bauthor{\bsnm{Yamada}, \binits{M.}}:
\batitle{Temporal intermittency in the energy cascade process and local
  lyapunov analysis in fully-developed model turbulence}.
\bjtitle{Progress of theoretical physics}
\bvolume{81}(\bissue{2}),
\bfpage{329}--\blpage{341}
(\byear{1989})
\doiurl{10.1143/PTP.81.329}
\end{barticle}
\endbibitem

%%% 3
\bibitem[\protect\citeauthoryear{Bohr et~al.}{1998}]{bohr1998dynamical}
\begin{bbook}
\bauthor{\bsnm{Bohr}, \binits{T.}},
\bauthor{\bsnm{Jensen}, \binits{M.H.}},
\bauthor{\bsnm{Paladin}, \binits{G.}},
\bauthor{\bsnm{Vulpiani}, \binits{A.}}:
\bbtitle{Dynamical Systems Approach to Turbulence}.
\bsertitle{Cambridge Nonlinear Science Series}.
\bpublisher{Cambridge University Press},
\blocation{New York}
(\byear{1998}).
\doiurl{10.1017/CBO9780511599972}
\end{bbook}
\endbibitem

%%% 4
\bibitem[\protect\citeauthoryear{Biferale}{2003}]{Biferale2003}
\begin{barticle}
\bauthor{\bsnm{Biferale}, \binits{L.}}:
\batitle{{Shell Models of Energy Cascade in Turbulence}}.
\bjtitle{Annual Review of Fluid Mechanics}
\bvolume{35}(\bissue{1}),
\bfpage{441}--\blpage{468}
(\byear{2003})
\doiurl{10.1146/annurev.fluid.35.101101.161122}
\end{barticle}
\endbibitem

%%% 5
\bibitem[\protect\citeauthoryear{Dombre and Gilson}{1998}]{Dombre1998}
\begin{barticle}
\bauthor{\bsnm{Dombre}, \binits{T.}},
\bauthor{\bsnm{Gilson}, \binits{J.L.}}:
\batitle{{Intermittency, chaos and singular fluctuations in the mixed
  Obukhov-Novikov shell model of turbulence}}.
\bjtitle{Physica D: Nonlinear Phenomena}
\bvolume{111}(\bissue{1-4}),
\bfpage{265}--\blpage{287}
(\byear{1998})
\doiurl{10.1016/S0167-2789(97)80015-2}
{\href{https://arxiv.org/abs/9510009}{{arXiv:9510009}}}
{[chao-dyn]}
\end{barticle}
\endbibitem

%%% 6
\bibitem[\protect\citeauthoryear{Mahrt}{1989}]{Mahrt89}
\begin{barticle}
\bauthor{\bsnm{Mahrt}, \binits{L.}}:
\batitle{Intermittency of atmospheric turbulence}.
\bjtitle{J. Atmos. Sci.}
\bvolume{46}(\bissue{1}),
\bfpage{79}--\blpage{95}
(\byear{1989})
\doiurl{10.1175/1520-0469(1989)046<0079:IOAT>2.0.CO;2"}
\end{barticle}
\endbibitem

%%% 7
\bibitem[\protect\citeauthoryear{Klymak et~al.}{2008}]{Rainville08}
\begin{barticle}
\bauthor{\bsnm{Klymak}, \binits{J.M.}},
\bauthor{\bsnm{Pinkel}, \binits{R.}},
\bauthor{\bsnm{Rainville}, \binits{L.}}:
\batitle{Direct breaking of the internal tide near topography: Kaena ridge,
  hawaii}.
\bjtitle{J. Phys. Oceanogr.}
\bvolume{38}(\bissue{2}),
\bfpage{380}--\blpage{399}
(\byear{2008})
\doiurl{10.1175/2007JPO3728.1}
\end{barticle}
\endbibitem

%%% 8
\bibitem[\protect\citeauthoryear{Falkovich and Pumir}{2008}]{Falkovich07}
\begin{barticle}
\bauthor{\bsnm{Falkovich}, \binits{G.}},
\bauthor{\bsnm{Pumir}, \binits{A.}}:
\batitle{Sling effect in collisions of water droplets in turbulent clouds}.
\bjtitle{J. Atmos. Sci.}
\bvolume{64}(\bissue{12}),
\bfpage{4497}--\blpage{4505}
(\byear{2008})
\doiurl{10.1175/2007JAS2371.1}
\end{barticle}
\endbibitem

%%% 9
\bibitem[\protect\citeauthoryear{Grafke et~al.}{2015}]{Grafke2015}
\begin{barticle}
\bauthor{\bsnm{Grafke}, \binits{T.}},
\bauthor{\bsnm{Grauer}, \binits{R.}},
\bauthor{\bsnm{Sch{\"{a}}fer}, \binits{T.}},
\bauthor{\bsnm{Vanden-Eijnden}, \binits{E.}}:
\batitle{{Relevance of instantons in Burgers turbulence}}.
\bjtitle{EPL (Europhysics Letters)}
\bvolume{109}(\bissue{3}),
\bfpage{34003}
(\byear{2015})
\doiurl{10.1209/0295-5075/109/34003}
\end{barticle}
\endbibitem

%%% 10
\bibitem[\protect\citeauthoryear{Margazoglou et~al.}{2019}]{Margazoglou2019}
\begin{barticle}
\bauthor{\bsnm{Margazoglou}, \binits{G.}},
\bauthor{\bsnm{Biferale}, \binits{L.}},
\bauthor{\bsnm{Grauer}, \binits{R.}},
\bauthor{\bsnm{Jansen}, \binits{K.}},
\bauthor{\bsnm{Mesterh{\'{a}}zy}, \binits{D.}},
\bauthor{\bsnm{Rosenow}, \binits{T.}},
\bauthor{\bsnm{Tripiccione}, \binits{R.}}:
\batitle{{Hybrid Monte Carlo algorithm for sampling rare events in space-time
  histories of stochastic fields}}.
\bjtitle{Physical Review E}
\bvolume{99}(\bissue{5}),
\bfpage{053303}
(\byear{2019})
\doiurl{10.1103/PhysRevE.99.053303}
{\href{https://arxiv.org/abs/1808.02020}{{arXiv:1808.02020}}}
\end{barticle}
\endbibitem

%%% 11
\bibitem[\protect\citeauthoryear{Apolin{\'a}rio
  et~al.}{2019}]{apolinario2019onset}
\begin{barticle}
\bauthor{\bsnm{Apolin{\'a}rio}, \binits{G.}},
\bauthor{\bsnm{Moriconi}, \binits{L.}},
\bauthor{\bsnm{Pereira}, \binits{R.}}:
\batitle{Onset of intermittency in stochastic burgers hydrodynamics}.
\bjtitle{Physical Review E}
\bvolume{99}(\bissue{3}),
\bfpage{033104}
(\byear{2019})
\doiurl{10.1103/PhysRevE.99.033104}
\end{barticle}
\endbibitem

%%% 12
\bibitem[\protect\citeauthoryear{Grafke et~al.}{2013}]{grafke2013instanton}
\begin{barticle}
\bauthor{\bsnm{Grafke}, \binits{T.}},
\bauthor{\bsnm{Grauer}, \binits{R.}},
\bauthor{\bsnm{Sch{\"a}fer}, \binits{T.}}:
\batitle{Instanton filtering for the stochastic burgers equation}.
\bjtitle{Journal of Physics A: Mathematical and Theoretical}
\bvolume{46}(\bissue{6}),
\bfpage{062002}
(\byear{2013})
\doiurl{10.1088/1751-8113/46/6/062002}
\end{barticle}
\endbibitem

%%% 13
\bibitem[\protect\citeauthoryear{Migdal}{2020}]{migdal2020clebsch}
\begin{barticle}
\bauthor{\bsnm{Migdal}, \binits{A.}}:
\batitle{Clebsch confinement and instantons in turbulence}.
\bjtitle{International Journal of Modern Physics A}
\bvolume{35}(\bissue{31}),
\bfpage{2030018}
(\byear{2020})
\doiurl{10.1142/S0217751X20300185}
\end{barticle}
\endbibitem

%%% 14
\bibitem[\protect\citeauthoryear{Moriconi and Pereira}{2022}]{moriconi2022}
\begin{barticle}
\bauthor{\bsnm{Moriconi}, \binits{L.}},
\bauthor{\bsnm{Pereira}, \binits{R.M.}}:
\batitle{Statistics of extreme turbulent circulation events from
  multifractality breaking}.
\bjtitle{Phys. Rev. E}
\bvolume{106},
\bfpage{054121}
(\byear{2022})
\doiurl{10.1103/PhysRevE.106.054121}
\end{barticle}
\endbibitem

%%% 15
\bibitem[\protect\citeauthoryear{Frick and Sokoloff}{1998}]{frick1998cascade}
\begin{barticle}
\bauthor{\bsnm{Frick}, \binits{P.}},
\bauthor{\bsnm{Sokoloff}, \binits{D.}}:
\batitle{Cascade and dynamo action in a shell model of magnetohydrodynamic
  turbulence}.
\bjtitle{Physical Review E}
\bvolume{57}(\bissue{4}),
\bfpage{4155}
(\byear{1998})
\doiurl{10.1103/PhysRevE.57.4155}
\end{barticle}
\endbibitem

%%% 16
\bibitem[\protect\citeauthoryear{Nigro et~al.}{2004}]{nigro2004nanoflares}
\begin{barticle}
\bauthor{\bsnm{Nigro}, \binits{G.}},
\bauthor{\bsnm{Malara}, \binits{F.}},
\bauthor{\bsnm{Carbone}, \binits{V.}},
\bauthor{\bsnm{Veltri}, \binits{P.}}:
\batitle{Nanoflares and mhd turbulence in coronal loops: a hybrid shell model}.
\bjtitle{Physical Review Letters}
\bvolume{92}(\bissue{19}),
\bfpage{194501}
(\byear{2004})
\doiurl{10.1103/PhysRevLett.92.194501}
\end{barticle}
\endbibitem

%%% 17
\bibitem[\protect\citeauthoryear{Plunian et~al.}{2013}]{PLUNIAN20131}
\begin{barticle}
\bauthor{\bsnm{Plunian}, \binits{F.}},
\bauthor{\bsnm{Stepanov}, \binits{R.}},
\bauthor{\bsnm{Frick}, \binits{P.}}:
\batitle{Shell models of magnetohydrodynamic turbulence}.
\bjtitle{Physics Reports}
\bvolume{523}(\bissue{1}),
\bfpage{1}--\blpage{60}
(\byear{2013})
\doiurl{10.1016/j.physrep.2012.09.001}
\end{barticle}
\endbibitem

%%% 18
\bibitem[\protect\citeauthoryear{Jensen et~al.}{1992}]{jensen1992shell}
\begin{barticle}
\bauthor{\bsnm{Jensen}, \binits{M.}},
\bauthor{\bsnm{Paladin}, \binits{G.}},
\bauthor{\bsnm{Vulpiani}, \binits{A.}}:
\batitle{Shell model for turbulent advection of passive-scalar fields}.
\bjtitle{Physical Review A}
\bvolume{45}(\bissue{10}),
\bfpage{7214}
(\byear{1992})
\doiurl{10.1103/PhysRevA.45.7214}
\end{barticle}
\endbibitem

%%% 19
\bibitem[\protect\citeauthoryear{Cohen et~al.}{2002}]{cohen2002statistically}
\begin{barticle}
\bauthor{\bsnm{Cohen}, \binits{Y.}},
\bauthor{\bsnm{Gilbert}, \binits{T.}},
\bauthor{\bsnm{Procaccia}, \binits{I.}}:
\batitle{Statistically preserved structures in shell models of passive scalar
  advection}.
\bjtitle{Physical Review E}
\bvolume{65}(\bissue{2}),
\bfpage{026314}
(\byear{2002})
\doiurl{10.1103/PhysRevE.65.026314}
\end{barticle}
\endbibitem

%%% 20
\bibitem[\protect\citeauthoryear{Mitra and Pandit}{2005}]{mitra2005dynamics}
\begin{barticle}
\bauthor{\bsnm{Mitra}, \binits{D.}},
\bauthor{\bsnm{Pandit}, \binits{R.}}:
\batitle{Dynamics of passive-scalar turbulence}.
\bjtitle{Physical review letters}
\bvolume{95}(\bissue{14}),
\bfpage{144501}
(\byear{2005})
\doiurl{10.1103/PhysRevLett.95.144501}
\end{barticle}
\endbibitem

%%% 21
\bibitem[\protect\citeauthoryear{Brandenburg}{1992}]{brandenburg1992energy}
\begin{barticle}
\bauthor{\bsnm{Brandenburg}, \binits{A.}}:
\batitle{Energy spectra in a model for convective turbulence}.
\bjtitle{Physical review letters}
\bvolume{69}(\bissue{4}),
\bfpage{605}
(\byear{1992})
\doiurl{10.1103/PhysRevLett.69.605}
\end{barticle}
\endbibitem

%%% 22
\bibitem[\protect\citeauthoryear{Ching}{2010}]{ching2010studying}
\begin{barticle}
\bauthor{\bsnm{Ching}, \binits{E.S.}}:
\batitle{Studying anomalous scaling and heat transport of turbulent thermal
  convection using a dynamical model}.
\bjtitle{Physica D: Nonlinear Phenomena}
\bvolume{239}(\bissue{14}),
\bfpage{1346}--\blpage{1352}
(\byear{2010})
\doiurl{10.1016/j.physd.2009.10.021}
\end{barticle}
\endbibitem

%%% 23
\bibitem[\protect\citeauthoryear{Campolina and
  Mailybaev}{2018}]{campolina2018chaotic}
\begin{barticle}
\bauthor{\bsnm{Campolina}, \binits{C.S.}},
\bauthor{\bsnm{Mailybaev}, \binits{A.A.}}:
\batitle{Chaotic blowup in the 3d incompressible euler equations on a
  logarithmic lattice}.
\bjtitle{Physical review letters}
\bvolume{121}(\bissue{6}),
\bfpage{064501}
(\byear{2018})
\end{barticle}
\endbibitem

%%% 24
\bibitem[\protect\citeauthoryear{Ditlevsen}{2011}]{Ditlevsen}
\begin{bbook}
\bauthor{\bsnm{Ditlevsen}, \binits{P.}}:
\bbtitle{Turbulence and Shell Models}.
\bpublisher{Cambridge University Press},
\blocation{New York}
(\byear{2011}).
\doiurl{10.1017/CBO9780511919251}
\end{bbook}
\endbibitem

%%% 25
\bibitem[\protect\citeauthoryear{D{\"{u}}ben et~al.}{2008}]{Duben2008}
\begin{barticle}
\bauthor{\bsnm{D{\"{u}}ben}, \binits{P.}},
\bauthor{\bsnm{Homeier}, \binits{D.}},
\bauthor{\bsnm{Jansen}, \binits{K.}},
\bauthor{\bsnm{Mesterhazy}, \binits{D.}},
\bauthor{\bsnm{M{\"{u}}nster}, \binits{G.}},
\bauthor{\bsnm{Urbach}, \binits{C.}}:
\batitle{{Monte Carlo simulations of the randomly forced Burgers equation}}.
\bjtitle{EPL (Europhysics Letters)}
\bvolume{84}(\bissue{4}),
\bfpage{40002}
(\byear{2008})
\doiurl{10.1209/0295-5075/84/40002}
\end{barticle}
\endbibitem

%%% 26
\bibitem[\protect\citeauthoryear{Farazmand and Sapsis}{2017}]{Farazmand2017}
\begin{barticle}
\bauthor{\bsnm{Farazmand}, \binits{M.}},
\bauthor{\bsnm{Sapsis}, \binits{T.P.}}:
\batitle{{A variational approach to probing extreme events in turbulent
  dynamical systems}}.
\bjtitle{Science Advances}
\bvolume{3}(\bissue{9}),
\bfpage{1701533}
(\byear{2017})
\doiurl{10.1126/sciadv.1701533}
\end{barticle}
\endbibitem

%%% 27
\bibitem[\protect\citeauthoryear{Mesterh{\'{a}}zy and
  Jansen}{2011}]{Mesterhazy2011}
\begin{barticle}
\bauthor{\bsnm{Mesterh{\'{a}}zy}, \binits{D.}},
\bauthor{\bsnm{Jansen}, \binits{K.}}:
\batitle{{Anomalous scaling in the random-force-driven Burgers' equation: a
  Monte Carlo study}}.
\bjtitle{New Journal of Physics}
\bvolume{13}(\bissue{10}),
\bfpage{103028}
(\byear{2011})
\doiurl{10.1088/1367-2630/13/10/103028}
\end{barticle}
\endbibitem

%%% 28
\bibitem[\protect\citeauthoryear{Schafer}{2004}]{Schafer:2004xa}
\begin{botherref}
\oauthor{\bsnm{Schafer}, \binits{T.}}:
{Instantons and Monte Carlo methods in quantum mechanics}
(2004)
\doiurl{10.48550/arXiv.hep-lat/0411010}
\end{botherref}
\endbibitem

%%% 29
\bibitem[\protect\citeauthoryear{Martin et~al.}{1973}]{Martin1973}
\begin{barticle}
\bauthor{\bsnm{Martin}, \binits{P.C.}},
\bauthor{\bsnm{Siggia}, \binits{E.D.}},
\bauthor{\bsnm{Rose}, \binits{H.A.}}:
\batitle{{Statistical Dynamics of Classical Systems}}.
\bjtitle{Physical Review A}
\bvolume{8}(\bissue{1}),
\bfpage{423}--\blpage{437}
(\byear{1973})
\doiurl{10.1103/PhysRevA.8.423}
\end{barticle}
\endbibitem

%%% 30
\bibitem[\protect\citeauthoryear{Janssen}{1976}]{Janssen1976}
\begin{barticle}
\bauthor{\bsnm{Janssen}, \binits{H.-K.}}:
\batitle{{On a Lagrangean for classical field dynamics and renormalization
  group calculations of dynamical critical properties}}.
\bjtitle{Zeitschrift für Physik B Condensed Matter and Quanta}
\bvolume{23}(\bissue{4}),
\bfpage{377}--\blpage{380}
(\byear{1976})
\doiurl{10.1007/BF01316547}
\end{barticle}
\endbibitem

%%% 31
\bibitem[\protect\citeauthoryear{Dominicis}{1976}]{dominicis1976technics}
\begin{bchapter}
\bauthor{\bsnm{Dominicis}, \binits{C.d.}}:
\bctitle{Technics of field renormalization and dynamics of critical phenomena}.
In: \bbtitle{J. Phys.(Paris), Colloq},
pp. \bfpage{1}--\blpage{247}
(\byear{1976}).
\doiurl{10.1051/jphyscol:1976138}
\end{bchapter}
\endbibitem

%%% 32
\bibitem[\protect\citeauthoryear{Onsager and Machlup}{1953}]{Onsager1953}
\begin{barticle}
\bauthor{\bsnm{Onsager}, \binits{L.}},
\bauthor{\bsnm{Machlup}, \binits{S.}}:
\batitle{{Fluctuations and Irreversible Processes}}.
\bjtitle{Physical Review}
\bvolume{91}(\bissue{6}),
\bfpage{1505}--\blpage{1512}
(\byear{1953})
\doiurl{10.1103/PhysRev.91.1505}
\end{barticle}
\endbibitem

%%% 33
\bibitem[\protect\citeauthoryear{Grafke et~al.}{2015}]{Grafke2015-2}
\begin{barticle}
\bauthor{\bsnm{Grafke}, \binits{T.}},
\bauthor{\bsnm{Grauer}, \binits{R.}},
\bauthor{\bsnm{Sch{\"{a}}fer}, \binits{T.}}:
\batitle{{The instanton method and its numerical implementation in fluid
  mechanics}}.
\bjtitle{Journal of Physics A: Mathematical and Theoretical}
\bvolume{48}(\bissue{33}),
\bfpage{333001}
(\byear{2015})
\doiurl{10.1088/1751-8113/48/33/333001}
\end{barticle}
\endbibitem

%%% 34
\bibitem[\protect\citeauthoryear{Stratonovich}{1957}]{stratonovich1957method}
\begin{bchapter}
\bauthor{\bsnm{Stratonovich}, \binits{R.}}:
\bctitle{On a method of calculating quantum distribution functions}.
In: \bbtitle{Soviet Physics Doklady},
vol. \bseriesno{2},
p. \bfpage{416}
(\byear{1957})
\end{bchapter}
\endbibitem

%%% 35
\bibitem[\protect\citeauthoryear{Hubbard}{1959}]{hubbard1959calculation}
\begin{barticle}
\bauthor{\bsnm{Hubbard}, \binits{J.}}:
\batitle{Calculation of partition functions}.
\bjtitle{Physical Review Letters}
\bvolume{3}(\bissue{2}),
\bfpage{77}
(\byear{1959})
\end{barticle}
\endbibitem

%%% 36
\bibitem[\protect\citeauthoryear{Duane et~al.}{1987}]{DUANE1987216}
\begin{barticle}
\bauthor{\bsnm{Duane}, \binits{S.}},
\bauthor{\bsnm{Kennedy}, \binits{A.D.}},
\bauthor{\bsnm{Pendleton}, \binits{B.J.}},
\bauthor{\bsnm{Roweth}, \binits{D.}}:
\batitle{Hybrid monte carlo}.
\bjtitle{Physics Letters B}
\bvolume{195}(\bissue{2}),
\bfpage{216}--\blpage{222}
(\byear{1987})
\doiurl{10.1016/0370-2693(87)91197-X}
\end{barticle}
\endbibitem

%%% 37
\bibitem[\protect\citeauthoryear{Lüscher}{2010}]{luscher2010computational}
\begin{botherref}
\oauthor{\bsnm{Lüscher}, \binits{M.}}:
Computational Strategies in Lattice QCD
(2010).
\doiurl{10.48550/arXiv.1002.4232}
\end{botherref}
\endbibitem

%%% 38
\bibitem[\protect\citeauthoryear{Brooks et~al.}{2011}]{MCMHandbook}
\begin{bbook}
\beditor{\bsnm{Brooks}, \binits{S.}},
\beditor{\bsnm{Gelman}, \binits{A.}},
\beditor{\bsnm{Jones}, \binits{G.}},
\beditor{\bsnm{Meng}, \binits{X.-L.}} (eds.):
\bbtitle{Handbook of Markov Chain Monte Carlo}.
\bpublisher{Chapman and Hall/{CRC}},
\blocation{Florida}
(\byear{2011}).
\doiurl{10.1201/b10905}
\end{bbook}
\endbibitem

%%% 39
\bibitem[\protect\citeauthoryear{Chernykh and Stepanov}{2001}]{Stepanov}
\begin{barticle}
\bauthor{\bsnm{Chernykh}, \binits{A.I.}},
\bauthor{\bsnm{Stepanov}, \binits{M.G.}}:
\batitle{Large negative velocity gradients in burgers turbulence}.
\bjtitle{Phys. Rev. E}
\bvolume{64},
\bfpage{026306}
(\byear{2001})
\doiurl{10.1103/PhysRevE.64.026306}
\end{barticle}
\endbibitem

%%% 40
\bibitem[\protect\citeauthoryear{Ebener et~al.}{2019}]{Ebener2018}
\begin{botherref}
\oauthor{\bsnm{Ebener}, \binits{L.}},
\oauthor{\bsnm{Margazoglou}, \binits{G.}},
\oauthor{\bsnm{Friedrich}, \binits{J.}},
\oauthor{\bsnm{Biferale}, \binits{L.}},
\oauthor{\bsnm{Grauer}, \binits{R.}}:
{Instanton based importance sampling for rare events in stochastic PDEs}.
Chaos
\textbf{29}
(2019)
\doiurl{10.1063/1.5085119}
\end{botherref}
\endbibitem

%%% 41
\bibitem[\protect\citeauthoryear{Schorlepp et~al.}{2021}]{Schorlepp_2021}
\begin{barticle}
\bauthor{\bsnm{Schorlepp}, \binits{T.}},
\bauthor{\bsnm{Grafke}, \binits{T.}},
\bauthor{\bsnm{Grauer}, \binits{R.}}:
\batitle{Gel'fand{\textendash}yaglom type equations for calculating
  fluctuations around instantons in stochastic systems}.
\bjtitle{Journal of Physics A: Mathematical and Theoretical}
\bvolume{54}(\bissue{23}),
\bfpage{235003}
(\byear{2021})
\doiurl{10.1088/1751-8121/abfb26}
\end{barticle}
\endbibitem

%%% 42
\bibitem[\protect\citeauthoryear{Moriconi et~al.}{2014}]{moriconi2014velocity}
\begin{barticle}
\bauthor{\bsnm{Moriconi}, \binits{L.}},
\bauthor{\bsnm{Pereira}, \binits{R.}},
\bauthor{\bsnm{Grigorio}, \binits{L.}}:
\batitle{Velocity-gradient probability distribution functions in a lagrangian
  model of turbulence}.
\bjtitle{Journal of Statistical Mechanics: Theory and Experiment}
\bvolume{2014}(\bissue{10}),
\bfpage{10015}
(\byear{2014})
\doiurl{10.1088/1742-5468/2014/10/P10015}
\end{barticle}
\endbibitem

%%% 43
\bibitem[\protect\citeauthoryear{Apolin{\'a}rio
  et~al.}{2019}]{apolinario2019instantons}
\begin{barticle}
\bauthor{\bsnm{Apolin{\'a}rio}, \binits{G.}},
\bauthor{\bsnm{Moriconi}, \binits{L.}},
\bauthor{\bsnm{Pereira}, \binits{R.}}:
\batitle{Instantons and fluctuations in a lagrangian model of turbulence}.
\bjtitle{Physica A: Statistical Mechanics and its Applications}
\bvolume{514},
\bfpage{741}--\blpage{757}
(\byear{2019})
\doiurl{10.1016/j.physa.2018.09.102}
\end{barticle}
\endbibitem

%%% 44
\bibitem[\protect\citeauthoryear{Yamada and Ohkitani}{1988}]{yamada1988goy}
\begin{barticle}
\bauthor{\bsnm{Yamada}, \binits{M.}},
\bauthor{\bsnm{Ohkitani}, \binits{K.}}:
\batitle{Lyapunov spectrum of a model of two-dimensional turbulence}.
\bjtitle{Phys. Rev. Lett.}
\bvolume{60},
\bfpage{983}--\blpage{986}
(\byear{1988})
\doiurl{10.1103/PhysRevLett.60.983}
\end{barticle}
\endbibitem

%%% 45
\bibitem[\protect\citeauthoryear{L'vov et~al.}{1998}]{lvov1998sabra}
\begin{barticle}
\bauthor{\bsnm{L'vov}, \binits{V.S.}},
\bauthor{\bsnm{Podivilov}, \binits{E.}},
\bauthor{\bsnm{Pomyalov}, \binits{A.}},
\bauthor{\bsnm{Procaccia}, \binits{I.}},
\bauthor{\bsnm{Vandembroucq}, \binits{D.}}:
\batitle{Improved shell model of turbulence}.
\bjtitle{Phys. Rev. E}
\bvolume{58},
\bfpage{1811}--\blpage{1822}
(\byear{1998})
\doiurl{10.1103/PhysRevE.58.1811}
\end{barticle}
\endbibitem

\end{thebibliography}

%\input sn-article.bbl

\end{document}